\documentclass[12pt,preprint]{aastex}
\usepackage{graphicx}

\usepackage{hyperref}

\usepackage{fancyhdr}
\usepackage[T1]{fontenc}
\usepackage[utf8]{inputenc}
\usepackage[svgnames]{xcolor}
\usepackage[paperheight=29.7cm,paperwidth=21.0cm,left=2.54cm,right=2.54cm,top=2.54cm,bottom=2.54cm]{geometry}
\usepackage{xurl}

\title{The Eclipse Megamovie Project (2017)}

\begin{document}
\maketitle

\vspace{2\baselineskip}
\begin{center}
\textbf{Hugh Hudson\\}
\textit{SSL UC Berkeley CA USA and University of Glasgow, UK\\
E-mail: hugh.hudson@glasgow.ac.uk}
\end{center}

\begin{center}
\textbf{Laura Peticolas\\}
\textit{Sonoma State University CA USA}
\end{center}

\begin{center}
\textbf{Calvin Johnson\\}
\textit{Google, Mountain View CA USA}
\end{center}

\begin{center}
\textbf{Vivian White\\}
\textit{Astronomical Society of the Pacific, San Francisco CA }
\end{center}

\begin{center}
\textbf{Mark Bender\\}
\textit{Eclipse Across America, San Antonio, TX USA}
\end{center}

\begin{center}
\textbf{Jay M. Pasachoff\\
Williams College-Hopkins Observatory,\\}
\textit{Williamstown MA 02167 USA}
\end{center}

\begin{center}
\textbf{Juan Carlos Mart{\' i}nez Oliveros\\}
\textit{Space Sciences Lab, UC Berkeley CA USA}
\end{center}

\begin{center}
\textbf{Alexei V. Filippenko\\}
\textit{UC Berkeley Astronomy Department, Berkeley CA USA}
\end{center}

\begin{center}
{\bf Christopher Cable, Braxton Collier, Noelle Filippenko, Andrew Fraknoi, Juan Camilo Guevara G{\' o}mez, Justin Koh, David Konerding, Larisza Krista, Brian Kruse, Scott McIntosh, Bryan Mendez, Igor Ruderman, Darlene Yan, Dan Zevin, and most importantly, many hundreds of volunteers}
\end{center}

{\footnotesize \textbf{\textcolor[HTML]{FFFFFF}{Abstract}}}

\noindent\textbf{Abstract:} The total solar eclipse of August 21, 2017, crossed the whole width of North America, the first occasion for this during the modern age of consumer electronics. Accordingly, it became a great opportunity to engage the public and to enlist volunteer observers with relatively high-level equipment; our program (``Eclipse Megamovie'') took advantage of this as a means of creating a first-ever public database of such eclipse photography. This resulted in a large outreach program, involving many hundreds of individuals, supported almost entirely on a volunteer basis and with the institutional help of Google, the Astronomical Society of the Pacific, and the University of California, Berkeley. The project home page is \url{http://eclipsemegamovie.org}, which contains the movie itself. We hope that our comments here will help with planning for similar activities in the total eclipse of April 8, 2024.

\section{INTRODUCTION}

\vspace{1\baselineskip}
This paper reviews a major public astronomy program (``The Eclipse Megamovie'') from the total solar eclipse of 2017 across North America. From a perspective halfway between this unique opportunity for crowd-sourced science, and the next one (2024), we discuss all aspects of the 2017 experience. Its uniqueness, briefly, reflects the rare geography involved (the whole breadth of North America could witness totality) plus the newly available consumer electronics at that epoch.

\vspace{1\baselineskip}
The basic idea for the Eclipse Megamovie program first surfaced in 2011, at a scientific meeting in Boulder, Colorado (a crossroads corner for solar physics in general). To get the ideas laid out we immediately wrote a descriptive ``white paper'' (Hudson \textit{et al.}, 2011). 
The scheme envisioned a major educational program for public outreach, plus a serious attempt to create a scientifically useful and public-domain record of the eclipse imagery. The basic science would consist of creating a huge (or at least very large) time-ordered array of images (the ``datacube''), containing as many image frames as possible, and therefore allowing searches at high time resolution for coronal dynamics. Viewing a datacube as a connected whole essentially means making a movie, and ideally a science-grade movie production, hence the name ``Eclipse Megamovie'' became program’s name and the title of its home page on the Web. From the outset we imagined getting a million or more still images all registered, scaled, and neatly assembled into one time-domain graphic (hence ``Megamovie''). Participation in this movie would reward the volunteers, whose names could possibly scroll through during the screening. In practice this program actually happened almost as intended, but with some flaws; see the Web site \url{http://eclipsemegamovie.org}. We missed the million mark but certainly acquired a large amount of excellent imagery. For reference, an estimated 154,000,000 individuals viewed this eclipse directly, according to one estimate, of whom about 0.001$\%$ became Megamovie volunteers contributing their own observations to the database, made with their own instruments and with the Megamovie program’s guidance and inspiration.

\vspace{1\baselineskip}
Eclipses have of course been inspirational and important in astronomy (and society) as a whole, and new science continues to come out of this eclipse observations of the solar corona, a fascinating part of the ``Plasma Universe.'' Major space missions (notably the current \textit{Parker Solar Probe} and the \textit{Solar Observer }spacecraft, both now in heliospheric orbit), continue the exploration of the corona, as do many Earth-based remote-sensing observations. (\textit{e.g.,} Paschoff $\&$ Fraknoi 2017).

\vspace{1\baselineskip}
We realized from the outset that the Megamovie datacube would not have much uniformity, as it would come from whatever good equipment a volunteer could bring to the event, as well as that volunteer's metadata (flat-fielding, dark frames \textit{etc.})
This left space for the Citizen CATE program, which actually provided standardized equipment and supervision for a smaller group of volunteer observers (62 of 68 returning data), also distributed across the North American continent (Penn \textit{et al.,} 2015). In principle these images, the Megamovie program's own database, and material from other observing groups could still be integrated into a single resource for this remarkable opportunity. This paper describes the outreach and the science ambitions of the Eclipse Megamovie, from a perspective several years after the fact but prior to a similar North American continental eclipse in 2024. To summarize: the Megamovie program succeeded well enough with data creation and archiving, but, perhaps expectably, the science results have not materialized. The heterogeneity of the database posed large problems, as did image coalignment -- this in spite of the fact that the Eclipse Megamovie guidance to observers encouraged them to include the bright star Regulus in the field of view of the recommended observing setup as an aid in image registration.
Only rarely do total eclipses of the Sun have such conveniently bright astrometric references!

\section{RESULTS: OUTREACH}

\vspace{1\baselineskip}
The public outreach portion of the program did outstandingly well before and at the time of the eclipse (Peticolas  \textit{et al}., 2018; White \textit{et al.,} 2018). The Eclipse Megamovie outreach program included an early buy of 2,000,000 pairs of eclipse glasses by Google; these went to schools, libraries, and other institutions along the path. Eclipse Megamovie team members also visited along the path before the event, lecturing and supervising ``town hall'' meetings, and of course we encouraged a presence as substantial as possible on social media and in print (Hudson \& Bender, 2017). The free app described below also reached many more people than could actually participate in the Megamovie observational program with full camera equipment.

\vspace{1\baselineskip}
Google sponsored a Website, now linked at \url{http://eclipsemegamovie.org}. This provided a registration point, and reached a total of more than 12,000 registrants. This site also linked an upload portal for volunteer image data and created an initial quick movie based on an image-sorting algorithm. 
Of these many registrants, the program accepted 1,190 to provide guidance and assistance as far as possible, limited in number at the maximum we felt we could properly deal with. Many of these participants succeeded well with the observations and most expressed pleasure about the opportunity to participate in the program. The quality of some of the volunteer imagery was quite excellent. The Astronomical Society of the Pacific supervised training for these people via author White, with tutors including veteran eclipse experts as well as experienced volunteers themselves; in practice a great deal of self-instruction augmented the training Webinar series. The program benefited by the helpful presence of Xavier Jubier, a great expert and innovator in the world of eclipse research. The volunteer observers each used their own equipment and operated within certain parameter ranges; for example we requested imaging fields of view large enough to include Regulus. We also requested, but did not succeed very well, the recording of flat fields and dark frames. The resulting scope of the program did not quite match the ``Mega'' level, which would have required a program more than an order of magnitude greater.

Based on participant commentary, we believe that Megamovie was a great success in terms of public outreach, especially in terms of awareness of the scientific implications of the eclipse phenomenon.

\section{RESULTS: SCIENCE}

\vspace{1\baselineskip}
The basic Eclipse Megamovie science lies in the characterization of time variability in the corona over the time of observation, taking advantage of the great length (some 90 minutes over land) of the 2017 eclipse. Longer time intervals would generally provide  more dramatic changes in the corona, but even at this scale we have distinct possibilities for novel research results. And, in fact, a slow CME flow occurred during the entirety of the eclipse period. To provide scale here, an easily recognizable image motion of 10’’ over 90 minutes corresponds to a speed of apparent motion of roughly 1 km/s; thus observations with this level of angular resolution could in principle detect and measure the ever-present solar rotation field.  Solar rotation already amounts to 2 km/s at the photosphere, and much of the structure of the lower corona links rigidly to the photosphere (but how this works physically remains largely unknown). 
Some structures must  \textit{co-rotate}, having larger velocities at higher altitudes, and becoming visible for foreground and background features not lying in the ``plane of the sky.'' Even 90 minutes of observation therefore could lead to some actual \textit{stereoscopic} information; note that the foreground and background features have opposite apparent motions (Li~\textit{et al.}, 2002). 
The 2017 eclipse in fact yielded a confirmation of the celebrated Eddington deflection of starlight 
(Dyson~\textit{et al.}, 1919), perhaps the first at visible wavelengths making use of modern digital detectors  (Bruns~\textit{et al.}, 2018).

\vspace{1\baselineskip}
Even by this time of writing, four years after the eclipse, we unfortunately have no new science results uniquely attributable to the Eclipse Megamovie program despite the advantages mentioned. The problem, as anticipated, lies in having few analysis resources confronting a very complicated database. A fully registered Megamovie was created by Berkeley co-authors Mart{\' i}nez Oliveros and Guevara G{\' o}mez, rescaling and rotating image via use of Regulus as an astrometric reference. This is a good guide to the breadth of the data (and is available on the Megamovie Web site) but contains no new science \textit{per se}. The 2017 eclipse was nevertheless particularly productive in several ways mentioned below.

\subsection{The stars in the field}

\vspace{1\baselineskip}
Much of the interest in eclipse imagery comes from the desire to make images with exquisite detail, for example in the beautiful results produced by Miloslav Druckmüller in many composited still images of eclipses\footnote{\url{http://www.zam.fme.vutbr.cz/~druck/Eclipse/index.htm}}. 
Extracting precise information from a time-ordered image datacube (a movie) has similar requirements, but they extend into the time domain. For the basic problem of image registration, fortunately, we often have the wonderful fiducial reference grid of the  background stars visible as they appear to drift through the eclipse image sequences. In 2017, most of the Megamovie images captured the first-magnitude star Regulus ($\alpha$ Leonis). Figure~1 shows eclipse observations of Regulus. Author I. Ruderman obtained these data with a KSON APO80-C telescope, 440 mm effective focal length at f/5.5$\}$, both the principal star and its faint companion Regulus B/C (located some 177$''$ away), itself a binary pair. In the 2$\sim$s exposures shown in the Figure, Regulus A badly saturates and cannot readily serve as a fiducial; indeed, in this set of images even the shortest exposures (1/6400~s in this sequence) show this bright star, giving more precise localization without the saturation problem. At least 13 stars appear in the sum of eight 2-s exposures shown in the Figure.  Such a long exposure saturates the inner corona quite badly (we cannot see into the annulus between 1 and 2 solar radii at this exposure time) and so we need a composite image to reveal the full dynamic range of the eclipse.

\begin{figure}[htbp]
\centering
   \includegraphics[width=0.8\textwidth]{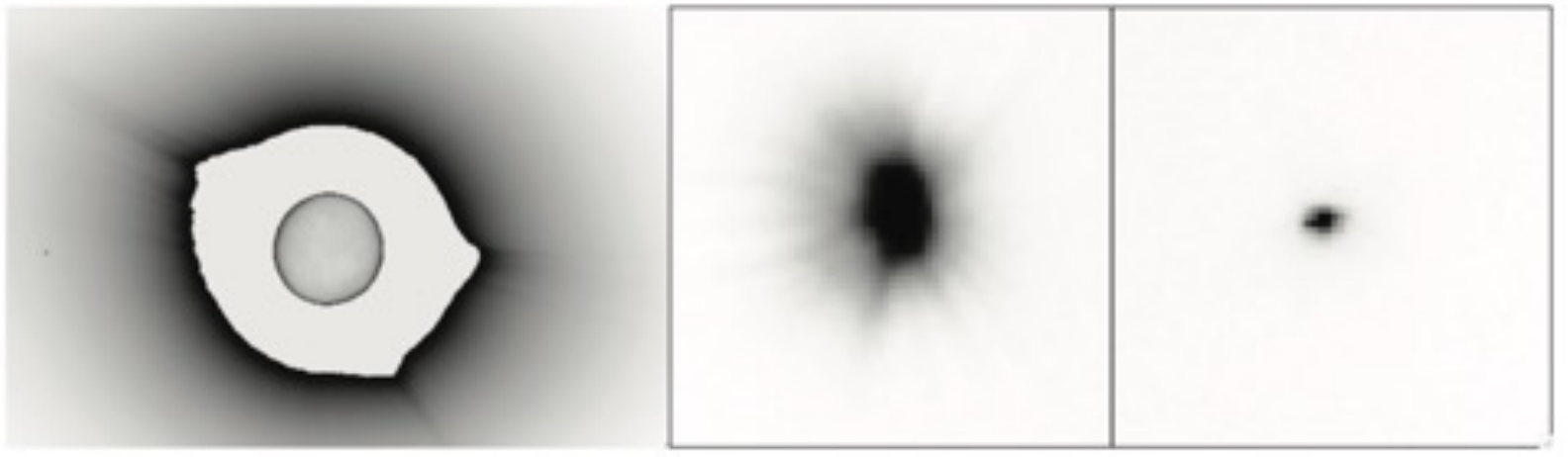}
           \caption{\textit{A 2-s image exposure of the 2017 eclipse field by author I. Ruderman, shown here in reversed color table with saturation set to white. Center, Regulus A (m\textsubscript{V} $=$ 1.3) and right, Regulus B/C (m\textsubscript{V} $=$ 8.15), as seen through the solar corona. These stellar images cover field of view of about 2’’ (64px); the Regulus$\sim$A image at this exposure time (2$\sim$s) has a high degree of saturation, and so one sees only the far wings of the scattering resulting from atmosphere and optics. But the Regulus B/C image looks compact and virtually unresolved. Based upon this we conclude that the stellar references in the best Eclipse Megamovie images could attain an alignment accuracy of order one arcsec. One can barely detect Regulus in the left-hand image, and only because its image has spread out so grossly because of the saturation: look at the same height as Moon center, near the left edge of the frame.. Note the possibly unfamiliar orientation here!} 
      }
\end{figure} 

\subsection{Image composites}

\vspace{1\baselineskip}
Most of the volunteers obtained eclipse images with brackets of exposures at different lengths, which they could then assemble into a ``high dynamic range'' (HDR) image to cover the whole range of intensities. The image brackets (sequences of exposures at varying duration) for the data in Figure 2, for example, spanned 1/6400~s to 2~s, a factor of some ten magnitudes or a factor 10\textsuperscript{4}; by contrast, a regular 8-bit image representation only covers a factor of 256 (10\textsuperscript{2.4}). Accordingly, to encompass the coronal structure across its large variation and to make it readily visible in a screen or printed graphic, one must do some compression.  We have combined this process with the multi-scale Gaussian normalization technique of \cite{2014SoPh..289.2945M} to produce Figure~2.

\begin{figure}[htbp]
\centering
   \includegraphics[width=0.6\textwidth]{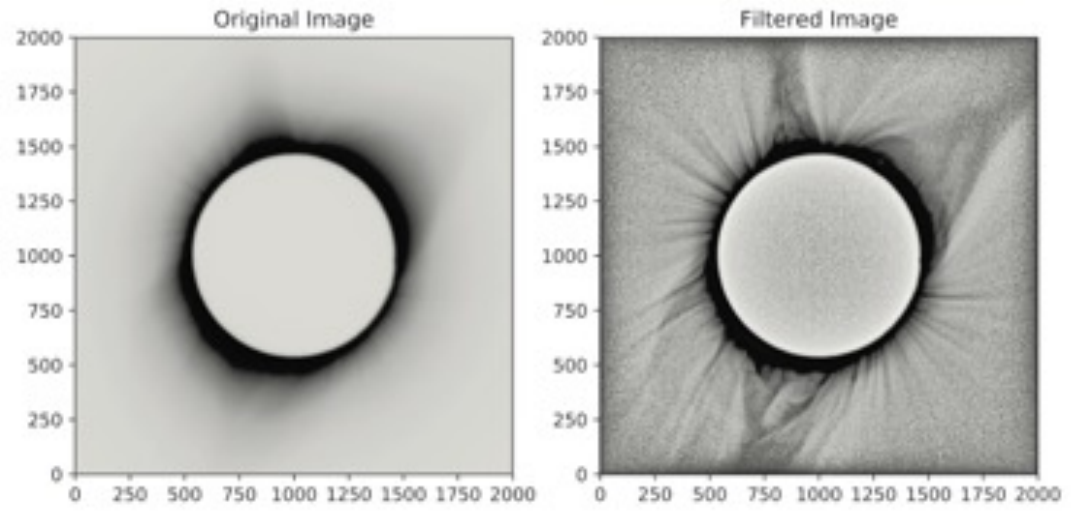}
           \caption{\textit{An illustration of the remarkable results of a conversion from an image bracket (several exposures of different times)  via the Morgan-Druckm{\" u}ller algorithm. The color table is reversed and the image orientation arbitrary in this case.} 
      }
\end{figure} 

Finally, image reconstructions as shown in Figure 2 have been registered, using Regulus as a reference, and strung together into a movie series. Figure 3 shows three consecutive HDR images, showing the stability of the process.

\begin{figure}[htbp]
\centering
   \includegraphics[width=0.6\textwidth]{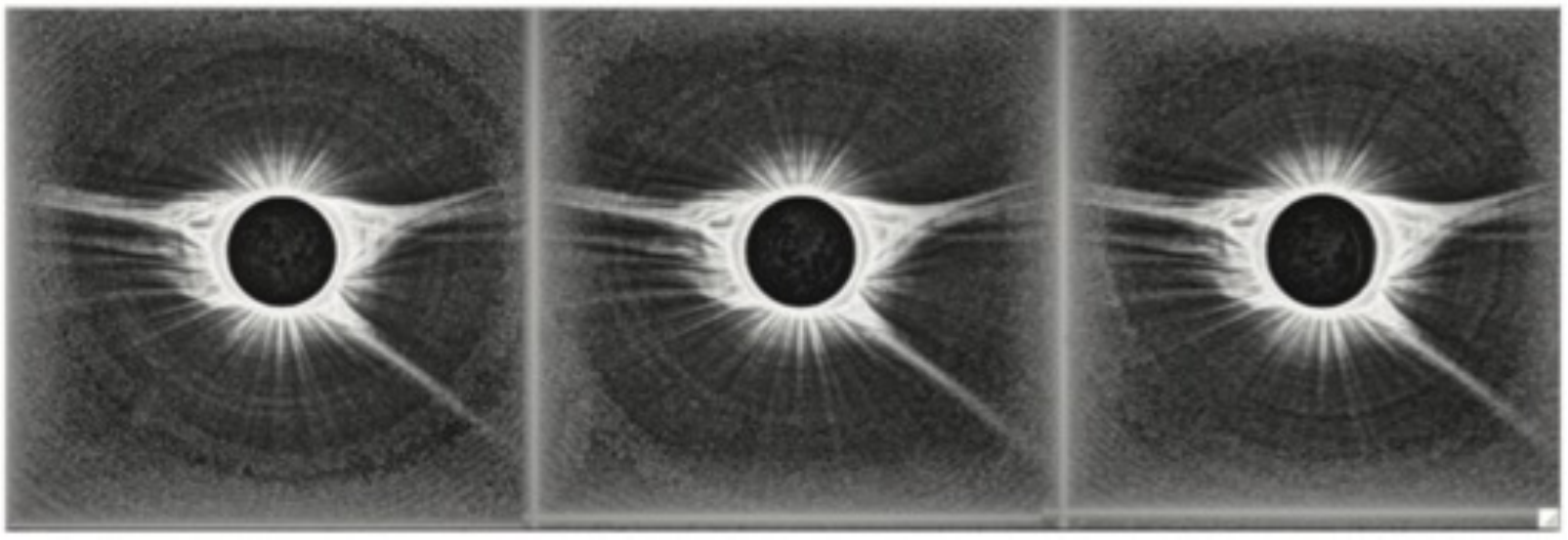}
           \caption{\textit{Three HDR images produced by the Morgan-Druckm{\" u}ller technique from one volunteer's bracketed image sets, and thus covering a time span of a few tens of seconds (positive color table). These are among the many HDR images of the database movie}.
      }
\end{figure} 

\vspace{1\baselineskip}
\subsection{Image results}

\vspace{1\baselineskip} 
Although we cannot report on Eclipse Megamovie discoveries of flows in the corona of August 21, 2017, other groups have already done so (\textit{e.g.} Pasachoff \textit{et al.} 2018; Hanaoka \textit{et al.} 2018), using equipment not appreciably different from that of some of the Eclipse Megamovie volunteers.
The latter made excellent observations of a specific kind of flow: collimated features, likely to be aligned on the magnetic field, associated with coronal jets. Hanaoka \textit{et al.}\ \ observe these jets as mass flows extending along open magnetic fields, and find that they match up with EUV/X-ray jets in the lower (\textit{e.g.} Shimojo \textit{et al.}, 1996). This confirms the jet development as a substantial and highly collimated material flow, rather than being the effect of wave propagation. The collimation seems inconsistent with the clearly observable and theoretically expectable divergence of the guiding magnetic field; why don't the jets expand laterally as they move radially?
 
\vspace{1\baselineskip}
We illustrate the Eclipse Megamovie archive's capability with a set of three HDR images, \textit{i.e.} ones composed of several exposures and sharpened with the Morgan-Druckmüller technique (Figure 2). Please see the Citizen CATE website (https://citizencate.org) for a movie composed of similar images but with a somewhat smaller field of view (Penn \textit{et al.}, 2020). Citizen CATE movies and Hanaoka \textit{et al. }difference images also captured the weak CME flow present during the eclipse time window.

\subsection{Status of data analysis}

The Eclipse Megamovie program has produced a mass of imagery, mostly without the careful attention to the flat-fielding and dark correction that basically convert images into actual scientific data. For future eclipses, notably the North American eclipse of April 8. 2024, it will be important to improve this aspect of the data acquisition. In addition to this problem, the full time series of imagery (the Megamovie itself) consists of data from a wide variety of sources, with different cameras, different observational schemes, different color responses, different image formats, and of course different atmospheric conditions for different observers. A detailed analysis \textit{could} reveal coronal dynamics with unprecedented oversampling (image cadence) and time resolution, but this would require considerable human effort.  With the imagery in a public-access archive, we can hope that individuals or groups, professional or amateur, will disentangle some of the best sequences (\textit{e.g.} Hanaoka~\textit{et al.}, 2016). We note also other successful programs of astronomical citizen-science imagery, such as \url{https://www.missionjuno.swri.edu/junocam/} or \url{Aurorasaurus (http://www.aurorasaurus.org}.

\section{A FREE APP FOR SMARTPHONE OBSERVATIONS}

The 2017 eclipse was the first continental-scale total eclipse for which we had wide\-spread availability of smartphones. These have many resources, including amazingly capable cameras and GPS connectivity. 
Aside from the main Megamovie activity with telescopes and DSLR cameras, the Megamovie program therefore provided a free app for smartphones that took advanrage of some of these resources, in addition to providing basic eclipse knowledge and training links. 
The Megamovie app computed precise timing predictions based on the user’s location, as obtained by GPS in real time. It also described and enabled the use of the smartphone camera with an accessory telescope. We found that an inexpensive 20$\times$ external lens would make rather good eclipse images. The app furthermore could execute camera control, freeing the user to enjoy the eclipse otherwise; and finally, the app could automatically upload the resulting image data to a Web repository. At a more elementary level, the app could simply do precise timing of the eclipse progress by using the built-in camera as a simple photometer. This allowed a user to record the precise timing (with precise GPS location) of the development of second and third contacts (the epochs of the Diamond Ring effect and Baily's Beads) in a completely hands-off manner.

Co-author Collier demonstrated the use of a smartphone camera with external telescopic optics during the eclipse. Figure 4 (left) shows an eclipse image taken through a 50x lens during the eclipse. Its quality is such that one can clearly see Regulus and, in principle, use it for image alignment. The right panel of the figure shows eight smartphones awaiting their program sequence, as defined by the Smartphone app, of imaging at second and third contacts.

\begin{figure}[htbp]
\centering
   \includegraphics[width=\textwidth]{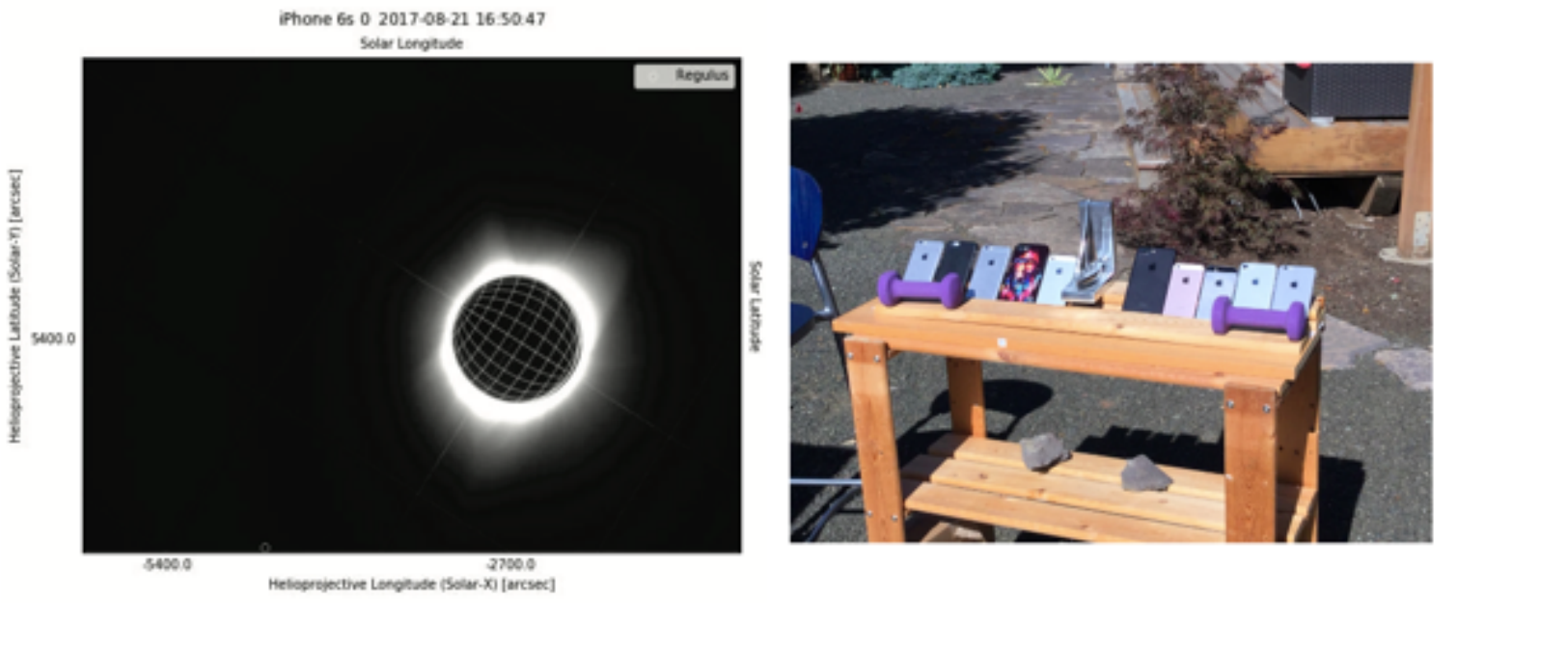}
           \caption{\textit{Left, an eclipse image captured by author Collier using a smartphone camera. Just barely visible at the bottom, with a small circle around it, is the image of Regulus (image preparation by author Mart{\' i}nez Oliveros). Right, a smartphone array at Corvallis, OR, between first and second contacts.}
      }
\end{figure} 

\vspace{1\baselineskip}
Why would we want to study the second and third contacts of an eclipse with a smartphone app?  First, of course, we thought that an image sequence showing the brief flash of the Diamond Ring effect would already please many smartphone owners and thus contribute to the outreach effort in science awareness. Second, the mass of precise timing data would contribute to an important scientific goal, already the subject of professional programs of eclipse photometry (\textit{e.g.} Lamy \textit{et al.} 2010). The shape of the Sun, at the highest resolution, departs from circularity in interesting and important ways, if only we could textit actually \textit{measure} them (\textit{e.g.}, Dicke~\textit{et al.}, 1974). At second contact, the bright arc of the solar disk rapidly collapses into one or more points of light created by lunar terrain -- the phenomenon of Baily's Beads. The  inverse effect, sudden point-like brightening, happens at third contact. These last and first gasps imply the possibility of extraordinarily high angular resolution for the determination of the position of the solar limb as a function of position angle (or heliographic latitude), given precise timing. 
Each user would thereby contribute two pieces of exact solar astrometry based on the lunar
reference. 
Within the last decade the \textit{Selene} and \textit{Lunar Reconnaissance Orbiter} probes have created a detailed 3D map of the Moon, accurate to within a few meters.  
For reference, 10 meters at the lunar distance corresponds to about 0.005$''$, 5 milliarcsec (MAS). 
Timing to one second corresponds to an angular motion of about 15$''$, so 1~msec corresponds to 15~MAS on an eclipse centerline, comparable in scale to our modern knowledge of the lunar surface structure. 

\section{ACCESS TO THE DATA}

The Eclipse Megamovie image archive now consists of 50,016 DSLR image frames obtained by the volunteer observers using their own DSLR/telescope equipment. The archive is publicly accessible via an SQL database at \url{https://goo.gl/J1zU7H}. This material contains full image metadata, except for personal identifications for the sake of security, as well as the actual images. The volunteer who provided the image brackets used in Figure 2, for example, had a unique 64-character identifer ending in \textit{6f4a73}, and an SQL search on that string should reveal relevant information such as the volunteer's location, the times of the images, the camera type, its technical settings, \textit{etc}. Shortly after the database went on line, one of the citizen scientists (Michael Andreas?) could readily make the Google map shown in Figure 5 by following the tag value ``totality.'' Generally, volunteer observers can readily find their contributions by precise location and time, via these SQL attributes.

\begin{figure}[htbp]
\centering
   \includegraphics[width=0.6\textwidth]{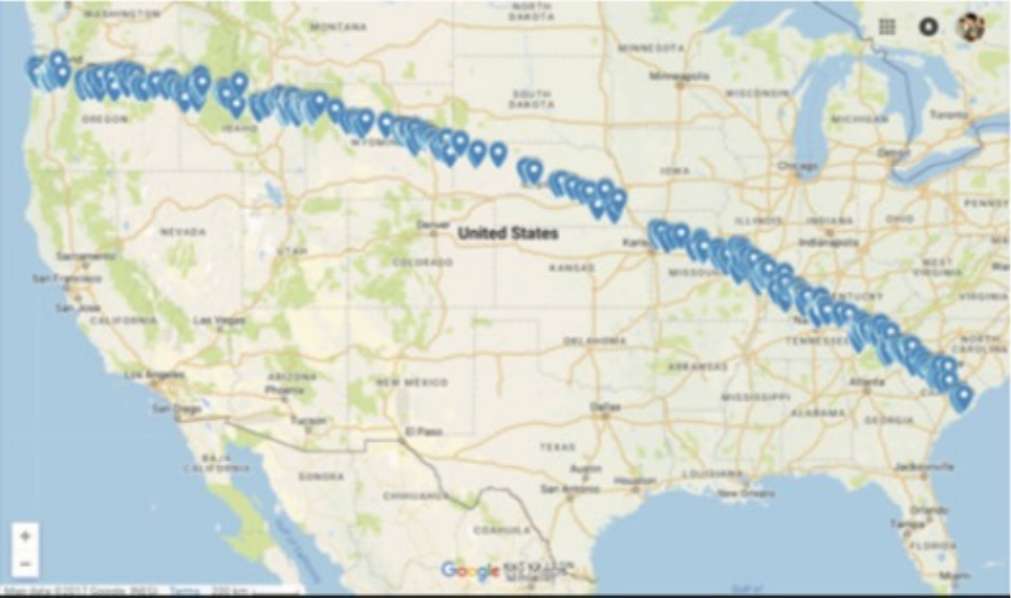}
           \caption{\textit{Megamovie volunteer locations during the eclipse, showing those with successful observations. Such a map was generated by one of the volunteers, using Google Maps and metadata from the public-domain Megamovie SQL database.}
      }
\end{figure} 

\section{CONCLUSIONS: WHAT ABOUT 2024?}

The Eclipse Eclipse Megamovie program accomplished most of its basic objectives: the outreach effort and the creation of the public database. The final objective, the identification of really new science, proved not so easy because of the complexity of the database, its limited volume and limited scientific manpower. We would love to have completed Megamovie by detecting flows and oscillations in the main DSLR program, and also to have been able to characterize the exact shape of the solar disk via the timing data obtained both from DSLR imagery and the smartphones, but did not meet these goals. In the meanwhile, the similar (but better funded) Citizen CATE activity did succeed quite well in stitching together images from many sites, and in demonstrating the scientific value of this eclipse movie approach (Penn \textit{et al}, 2020; see also Hanaoka~\textit{et al.} 2018). 

Based on the experience in 2017 and in the continuing program, we see things that we can do better at the next opportunity. We dream that a similar program will capture the major North American total eclipse of 2024 (See Xavier Jubier's map\footnote{http://xjubier.free.fr/en/site\_pages/solar\_eclipses/TSE\_2024\_GoogleMapFull.html} for example, when another and even better opportunity will occur).  This will cross the continent again, this time starting in Mazatl{\' a}n, Mexico, crossing Dallas/Fort Worth, Indianapolis, and Cleveland among big cities in the US, then exiting via Montr{\' e}al and Gander (Newfoundland) in Canada. We should regard the 2017 experience as a trial run, with a participation that only amounted to a tiny fraction of one percent of the potential observer base. The dream for 2024 envisions orders-of-magnitude increases in the volume of data archived and studied.

A comparable 2024 program would be highly desirable from the outreach point of view. Our advice, based upon the Megamovie experience, would be to be sure of some continuity in funding support for the program. We have not discussed future technical matters here, but note that observations of the ``flash spectrum'' of the chromosphere are well within community capabilities. On the data analysis side, the most important item would be the follow-up analysis. We speculate that a future program may make use of the rapidly growing machine-learning capabilities in ``big data'' applications.

We take this opportunity to mention some of the obstacles in the 2017 experience. First, the program failed to attract support from the major US funding agencies (NSF and NASA), and furthermore was not even welcomed at no cost in the huge public-outreach activity that this eclipse engendered. Possibly this resulted from different objectives: many public outreach activities strive to \textit{push} information to everybody; Megamovie basically wished to \textit{pull} data from them as well. Second, it is difficult to motivate programs of this type on the basis of actual science prospects; naturally a precisely tailored professional program is likeliest to be successful, and peer review must recognize this -- it is harder to justify an interdisciplinary program. In the last analysis, the eager participants in Megamovie observations could not provide hands-on data analysis, due in large part to the difficulty of managing a large program with no funding!

Finally, we must report substantial disappointment with the performance of the smartphone app. Although it functioned well, and was widely downloaded, almost no users succeeded in eclipse imagery with external lenses. The ``smartphone as a photometer" capability did not work well, although some data did appear on the Web portal as planned. Part of the problem was simply that the Megamovie app did not succeed in gaining a large market share; other apps with much less sophistication were available and were more popular. It may also be that many eclipse-watchers preferred to use their smartphone cameras for other purposes during the eclipse, for example to take selfies. Finally a concept flaw in the exposure programming reduced the amount of data, and in the final tally there was not enough timing information to warrant scientific analysis at a fundamentally interesting level. Some of these problems could be overcome in a 2024 version.

Based on our success in Megamovie (and in similar activities in 2017) we think that a similar program should follow in 2024. It should be still grander in scope and have proper support from sponsors. The main lesson from Megamovie was that a program of this nature, involving thousands of participants, definitely requires a professional (ie, well-supported) multidisciplinary management team.

\section{AUTHOR CONTRIBUTIONS}

HH wrote the initial draft of this paper, with help from AF, and contributed general science overview and input to the app development; LP managed the overall Eclipse Megamovie program, and contributed to the data analysis, CJ led the effort at Google, and contributed to the data analysis; VW led the program at the Astronomical Society of the Pacific, including the instructional Webinars; MB co-authored our \textit{Sky $\&$ Telescope} article, worked with the app development, produced instructional videos and other media efforts, tested the app in Patagonia in the next earlier eclipse, and helped with the data analysis; BC wrote the app and tested smartphone camera astronomy; BK worked with the outreach effort; SM contributed to data analysis after his inspirational contribution to the original idea; JCMO helped detector development, observations, and with data reductions, including the Morgan-Druckmüller\ \ image processing (with JCGG); JMP contributed valuable practical and theoretical advice, and helped with data analysis; IR contributed to data analysis and also acquired the images used in Figure 1, and DM worked with outreach and archiving.

\noindent{\bf Acknowledgements: }We thank Google, the Astronomical Society of the Pacific, and the Space Sciences Laboratory at the University of California, Berkeley, for support, and we hugely thank the Eclipse Megamovie volunteers. We also thank Xavier Jubier for substantial help and for creating his excellent ``Eclipse Maestro'' free software.

\nocite{2016SASS...35...49B}
\nocite{1974ApJS...27..131D}
\nocite{1920RSPTA.220..291D}
\nocite{2008Sci...322..560F}
\nocite{2018ApJ...860..142H}
\nocite{2011arXiv1108.3486H}
\nocite{2015SoPh..290.2617}
\nocite{2002ApJ...565.1289L}
\nocite{2014SoPh..289.2945M}
\nocite{2017AAS...23010804P}
\nocite{2018FrASS...5...37P}
\nocite{2015SASS...34...63P}
\nocite{2017PASP..129a5005P}
\nocite{2019ASPC..516..337P}
\nocite{1996PASJ...48..123S}
\nocite{2018...W}

\bibliographystyle{apj}
\bibliography{arxiv.bib}        

\end{document}